\documentclass[aip,jcp,reprint]{revtex4-1}
\usepackage[utf8]{inputenc}
\usepackage[T1]{fontenc}
\usepackage{ae}
\usepackage{amsbsy}
\usepackage{amstext}
\usepackage{amssymb}
\usepackage{graphicx}
\usepackage{MnSymbol}
\usepackage{wasysym}
\usepackage{mathdots}
\usepackage{esint}
\usepackage{url}
\usepackage{amsmath}
\usepackage{cprotect}

\usepackage{soul,xcolor} 
\setstcolor{red}

\makeatletter

\@ifundefined{textcolor}{}
{%
 \definecolor{BLACK}{gray}{0}
 \definecolor{WHITE}{gray}{1}
 \definecolor{RED}{rgb}{1,0,0}
 \definecolor{GREEN}{rgb}{0,1,0}
 \definecolor{BLUE}{rgb}{0,0,1}
 \definecolor{CYAN}{cmyk}{1,0,0,0}
 \definecolor{MAGENTA}{cmyk}{0,1,0,0}
 \definecolor{YELLOW}{cmyk}{0,0,1,0}
 }

\usepackage{bbm}
\usepackage[english]{babel}

\makeatother

\begin{document}

\title{Nonequilibrium Processes in Repulsive Binary Mixtures}

\author{Pedro Antonio Santos-Flórez}
\affiliation{Instituto de Física "Gleb Wataghin", Universidade Estadual de Campinas, UNICAMP, 13083-859, Campinas, São Paulo, Brazil}

\author{Maurice de Koning}
\email{dekoning@ifi.unicamp.br}
\affiliation{Instituto de Física "Gleb Wataghin", Universidade Estadual de Campinas, UNICAMP, 13083-859, Campinas, São Paulo, Brazil}
\affiliation{Center for Computing in Engineering \& Sciences, Universidade Estadual de Campinas, UNICAMP, 13083-861, Campinas, São Paulo, Brazil}

\date{\today}
\begin{abstract}

We consider rapid cooling processes in classical, 3-dimensional, purely repulsive binary mixtures in which an initial infinite-temperature (ideal-gas)  configuration is instantly quenched to zero temperature. It is found that such systems display two kinds of ordering processes, the type of which can be controlled by tuning the interactions between unlike particles. While strong inter-species repulsion leads to chemical ordering in terms of an unmixing process, weak repulsion gives rise to spontaneous crystallization, maintaining chemical homogeneity. This result indicates the existence of a transition in the topography of the underlying potential-energy landscape as the intra-species interaction strength is varied. Furthermore, the dual-type behavior appears to be universal for repulsive pair-interaction potential-energy functions in general, with the propensity for the crystallization process being related to their behavior in the neighborhood of zero separation.
\end{abstract}
\maketitle

\section{Introduction}

Classical systems described by repulsive pair potentials have been the subject of intense investigation for over five  decades.~\cite{Stillinger1964,Helfand1968,Baram1983,Baram1990,Baram1991,Baram1992,Baram1993,Dijkstra1998,Lang2000,Louis2000,Speedy2003,Cinacchi2007,Glaser2007,Saija2009,Berthier2010a,Schmiedeberg2011,Russo2012a,Travesset2015,Horst2016,Stillinger1976,Hansen2000,Prestipino2005,Prestipino2005a,Malescio2003,Mladek2006,Likos2007,Likos2008,Overduin2009,Archer2004,Shin2009,Shall2010,Russo2012a,Travesset2015,Horst2016} Not in the least due to their role as effective descriptions for interactions in soft-condensed-matter systems,~\cite{Likos1998,Watzlawek1999,Ferber2000,Likos2001,Likos2002,Likos2006,Mayer2008,Mladek2008,Mayer2009,Likos2011,Nikoubashman2015} substantial effort has been directed towards elucidating the equilibrium phase behavior of such models, considering both single-component samples as well as multi-component mixtures.~\cite{Stillinger1976,Hansen2000,Prestipino2005,Prestipino2005a,Malescio2003,Mladek2006,Likos2007,Likos2008,Overduin2009,Archer2004,Shin2009,Shall2010,Russo2012a,Travesset2015,Horst2016} 

Nonequilibrium phenomena, on the other hand, have received much less attention, despite their key role in self-organization phenomena in such systems.~\cite{Nicolis1977,Witten1999,Marson2014,Kalsin2006,Miller2009,Ye2015,Kumar2017,Ye2011,Macfarlane2011,Knorowski2011,Knorowski2011a} Indeed, one of the challenges in soft-matter materials design concerns the ability to adjust the effective interaction parameters so as to control the self-organization process and achieve desired self-assembled structures.~\cite{Knorowski2011a} In this context, processes that display spontaneous development of structure from an initially disordered, far-from-equilibrium state are of particular interest.~\cite{Knorowski2011,Knorowski2011a,Nicolis1977} 

A typical example is a process in which a system initially at equilibrium in a high-temperature state is suddenly quenched to low temperature.~\cite{Henkel2011} Because of the instant cooling, the initial high-temperature phase  falls out of equilibrium and spontaneously decays into a low-temperature state.   However, due to the intrinsic nonequilibrium nature of this cooling protocol the corresponding low-temperature state most often does not correspond to that given by the equilibrium phase diagram, characterizing the intrinsic nonequilibrium nature of the process.

 When considering mixtures, the sudden quench of a high-temperature state can give rise to two kinds of decay processes.~\cite{Balluffi2005,Henkel2011}  The prototypical example of the first kind are unmixing phenomena in which the final low-temperature state is characterized by chemical ordering through phase separation, whereas the second type is typified by the development of structural order. While unmixing transitions are quite common for the class of repulsive pair potentials,~\cite{Archer2001,Likos2011,Kambayashi1992} the occurrence of the second type of process is not. In fact, as far as model systems are concerned, to the best of our knowledge such structural ordering phenomena have so far only been observed for discrete spin systems such as the Ising model,~\cite{Henkel2011} while there have been no reports for systems characterized by continuous interactions. Above all, to date there are no known model systems that can display both types of processes as a function of boundary conditions and/or model parameters. 

Here, we show that 3-dimensional binary mixtures described by purely repulsive pairwise interactions display both kinds of  decay processes and that the observed type can be controlled by tuning the interactions between unlike particles. While strong inter-species repulsion gives rise to chemical ordering through unmixing, weak values lead to a spontaneous development of structural order, forming a polycrystalline solid of uniform chemical composition. Interestingly, this crystallization process is barrierless in nature and gives rise to grain-size distributions that display scale-invariant characteristics. Furthermore, the results suggest that the dual-type decay behavior is universal for pairwise repulsive potential-energy functions in general and that the propensity of the crystallization process is related to their behavior in the neighborhood of zero separation. 

\section{Computational Approach}

\subsection{Simulation Protocol}
\label{protocol}

We consider the case in which the cooling process is infinitely rapid, meaning that the initial infinite-temperature (i.e., ideal-gas) state is instantly quenched to zero temperature. Because the quench is infinitely fast, the system has no time to explore the potential-energy landscape (PEL) and is instantaneously driven to the local minimum closest to the initial configuration, also known as its inherent structure (IS).~\cite{Stillinger1995,Stillinger2015,Wales2003} In this sense, the process is fundamentally different from a quasi-static cooling protocol in which the system is at equilibrium at all times and the outcome is determined by the equilibrium phase diagram. 

The quench process is implemented computationally in the following way. First, for a specified particle density, we construct a cubic, periodic simulation cell with a volume $V$ that corresponds to a given total particle number $N$. Subsequently, the system is initialized by randomly placing the $N$ particles in the cell, giving rise to a structureless, uniform position distribution that represents an infinite-temperature, i.e., ideal-gas state. Then, to locate the corresponding IS, a conjugate-gradient (CG) minimization is invoked. For each set of interaction properties and particle densities this procedure is repeated several times using different random initial conditions. All the CG calculations have been performed using the Polak-Ribiere version of the CG algorithm as implemented in the \verb|LAMMPS| package,~\cite{Plimpton1995} which is among the most efficient local minimization algorithms for functions of many variables.~\cite{Press2007}

\subsection{Interaction Models} 
\label{models}

\begin{figure} [ht!]
	\includegraphics[width=8.5 cm ]{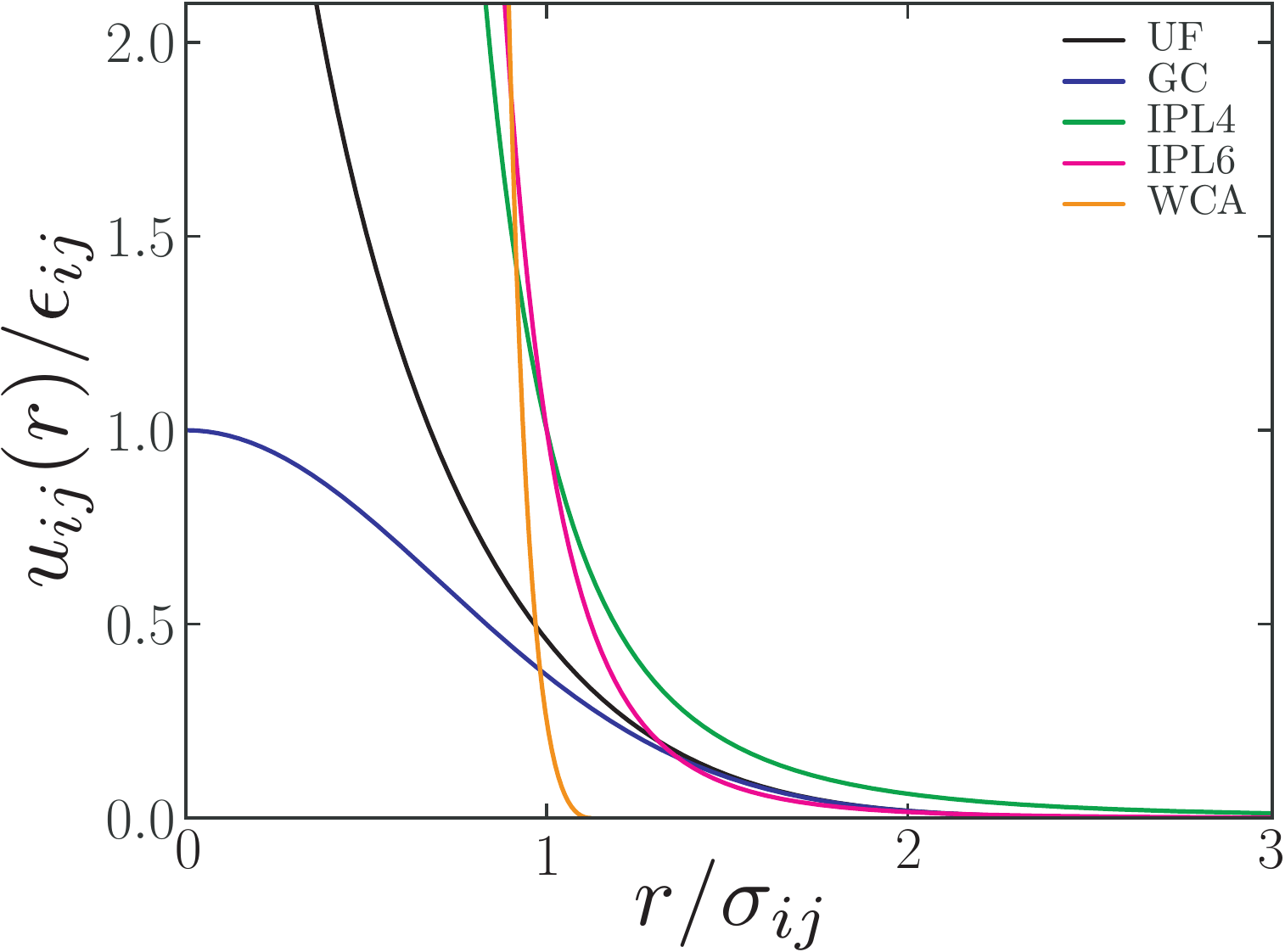}
	\caption{(Color online) Graphs of considered repulsive pair potentials. }
	\label{Fig1}
\end{figure}
We carry out the quench protocols described in Sec.~\ref{protocol} for a set of five different purely repulsive pair interaction models. Specifically, we consider, (i), the Uhlenbeck-Ford (UF) model,~\cite{deBoer1962,PaulaLeite2016,PaulaLeite2017,PaulaLeite2019} (ii), the inverse fourth (IPL4) and sixth-power (IPL6) potentials,~\cite{Kambayashi1992}, (iii), the Weeks-Chandler-Andersen (WCA) force field,~\cite{Weeks1971} and, (iv), the Gaussian-core (GC) potential.~\cite{Stillinger1976}

The UF model~\cite{deBoer1962,PaulaLeite2016,PaulaLeite2017,PaulaLeite2019} is  characterized by a logarithmic divergence at zero separation and belongs to the class of so-called ultrasoft potentials.~\cite{Likos2002} It is defined by the potential-energy function 
\begin{equation}
u_{ij}(r)=-\epsilon_{ij}\ln(1-e^{-r^2/\sigma_{ij}^2}),
\end{equation}
where $r$ is the interparticle distance, the indices $i$ and $j$ denote the species of the interacting particles (either $A$ or $B$) and $\epsilon_{ij}$ and $\sigma_{ij}$ are the corresponding energy and length scales. 
The cut-off for the interaction calculation is set at $r_c=4\,\sigma$.

The potential-energy functions describing the IPL4 and IPL6 models~\cite{Kambayashi1992} are given by 
\begin{equation}
u_{ij}(r)=\epsilon_{ij}\left(\frac{\sigma_{ij}}{r}\right)^n,
\end{equation}
The exponents for the IPL4 and IPL6 models are $n=4$ and $n=6$, respectively, and the cut-offs for the interaction calculation are set at $r_c=6\,\sigma$ and $r_c=4\,\sigma$ for the IPL4 and IPL6 models.

The WCA model~\cite{Weeks1971} is defined by the repulsive part of the Lennard-Jones (LJ) potential energy function, shifting the LJ function such that the minimum value corresponds to zero, and truncating it for distances beyond that of its minimum at $r=2^{1/6}\,\sigma$. Accordingly, the WCA force field is defined as
\begin{equation}
    u_{ij}(r)= 
\begin{cases}
    4\, \epsilon_{ij}\left[\left(\frac{\sigma_{ij}}{r}\right)^{12}-\left(\frac{\sigma_{ij}}{r}\right)^{6}\right]+\epsilon_{ij},& \text{if } r\leq 2^{1/6}\, \sigma_{ij} \\
    0,              & \text{otherwise.}
\end{cases}
\end{equation}

Finally, the GC model also belongs to the category of ultrasoft interaction models and is defined as~\cite{Stillinger1976}
\begin{equation}
u_{ij}(r)=\epsilon_{ij} \exp(-r^2/\sigma^2_{ij}).
\end{equation}

Figure~\ref{Fig1} compares the behaviors of these interaction models, plotting $u/\epsilon_{ij}$ as a function of the scaled interparticle distance $r/\sigma_{ij}$. The main difference between these models is their behavior near the origin. The UF, IPL4, IPL6 and WCA models all diverge at the origin, yet at different rates. Whereas the UF model diverges only logarithmically, the IPL4, IPL6 and WCA force fields diverge according to the inverse powers $r^{-4}$, $r^{-6}$ and $r^{-12}$, respectively. The GC, on the other hand, does not diverge at all as $r\to 0$, reaching a constant value at zero slope. 

\begin{figure*} [ht!]
	\includegraphics[width=17 cm ]{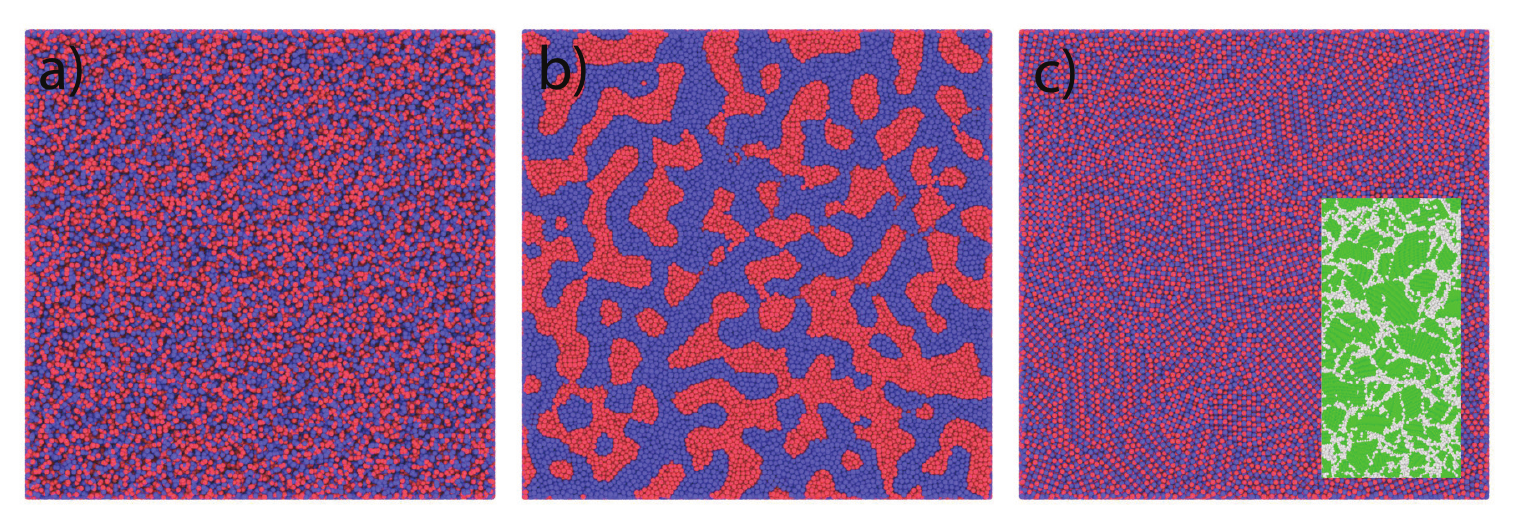}
	\cprotect\caption{(Color online) Typical configurations containing $N=10^7$ particles, with the distinct species shown in blue and red, respectively, as obtained from the CG quench protocol for the binary UF model for two different values for the inter-species interaction energy scale. a) Infinite-temperature (i.e., ideal-gas) initial configuration. b) Phase-separating system for strong inter-species repulsion. c) Spontaneous ordering into a rock-salt (B1) type polycrystal of uniform composition. Inset displays part of the grain structure, with the green and white particles representing those in the B1 structure and in the disordered surroundings of the grain boundaries, respectively, as determined using the \verb|Ovito| package.~\cite{Stukowski2010a,Larsen2016a}}
	\label{Fig2}
\end{figure*}
\section{Results and Discussion}

For all models we fix the energy scales of the interactions between particles of the same species to be $\epsilon_{AA}=100 \, \epsilon$ and $\epsilon_{BB}=200 \,\epsilon$, respectively, whereas $\epsilon_{AB}$ for inter-species interactions between $A$ and $B$ particles is variable. The length scale is chosen to be the same for all interaction types, i.e., $\sigma_{AA}=\sigma_{BB}=\sigma_{AB}=\sigma$. Furthermore, we consider symmetric binary mixtures in all cases, meaning that species $A$ and $B$ are present in equal proportions.

As a first case we consider  the quench-simulation results for the binary mixture described by UF inter-particle  model.~\cite{deBoer1962,PaulaLeite2016,PaulaLeite2017,PaulaLeite2019} Figure~\ref{Fig2} displays typical configurations obtained for the UF mixture containing $10^7$ particles at a reduced particle density $\rho^*\equiv N\sigma^3/V=1$. Figure~\ref{Fig2}a) depicts a typical random ideal-gas initial condition that is completely disordered, both chemically and structurally. Figures~\ref{Fig2}b) and c) then show snapshots obtained from the subsequent CG minimizations for two different values of the inter-species interaction parameter, $\epsilon_{AB}$. 

Fig.~\ref{Fig2}b) portrays a case of strong inter-species repulsion at $\epsilon_{AB}=175 \epsilon$. Under these conditions the initial ideal-gas phase undergoes a chemical ordering transition by which the two species unmix.  Indeed, the depicted structure strongly resembles the typical patterns of spinodal decomposition often seen for phase separation.~\cite{Balluffi2005,Laradji1996,Thakre2008} Note, however, that the structure depicted in Fig.~\ref{Fig2}b) has not yet fully converged to the completely unmixed IS. This is because the computational cost to reach a fully unmixed state is prohibitively large for the system size considered here, even for efficient minimizers such as CG. For smaller system sizes, however ($N\sim 10^5-10^6$), complete unmixing is attained within reasonable computational limits.

For a weak inter-species interaction at $\epsilon_{AB}=20 \epsilon$, the decay phenomenon is fundamentally different. In this case the CG minimization rapidly converges to the IS displayed in Fig.~\ref{Fig2}c), which remains uniform with respect to chemical composition but has spontaneously developed structural order. In particular, it features a polycrystalline morphology composed of grains with the rock-salt (B1) structure, which consists of two interpenetrating fcc lattices, each occupied by either $A$ or $B$. Interestingly, the nature of this crystallization process is rather different from the usual equilibrium freezing phenomena, which occur by nucleation and growth. Here, the crystallization transition  between the  initial ideal-gas  phase and the final polycrystalline structure is barrierless since they are connected by a CG sequence that always moves downhill on the PES.~\cite{Press2007} 

\begin{figure} [ht!]
	\includegraphics[width=8.5 cm ]{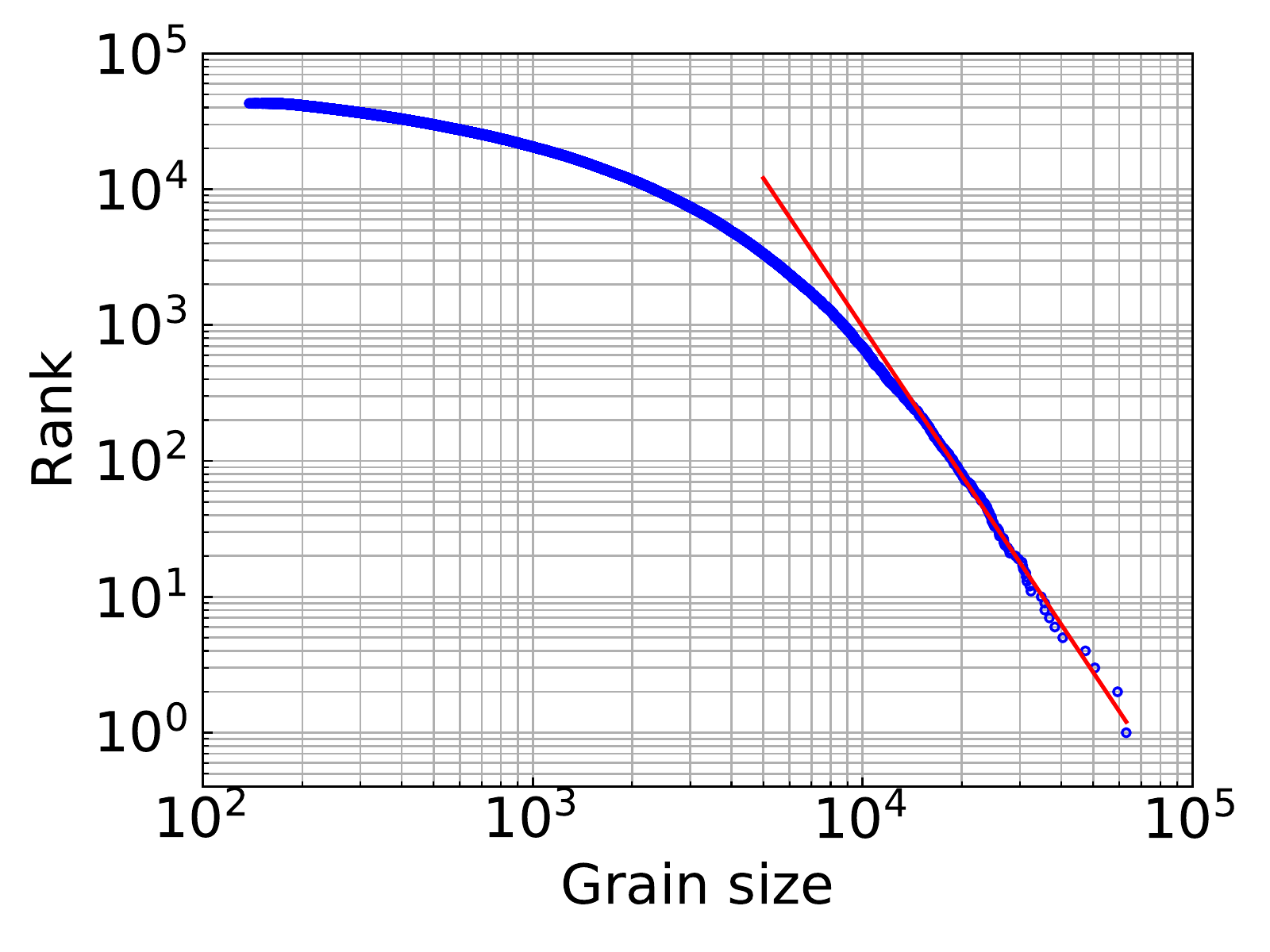}
	\cprotect\caption{(Color online) Log-log graph of rank-size representation of the grain size distribution for cell containing $10^8$ particles as obtained using the grain-segmentation tool of the \verb|Ovito| package~\cite{Stukowski2010a,Larsen2016a} , plotting the rank of each grain as a function of its size $k$ such that the largest and smallest grains are ranked first and last, respectively. Blue circles depict results data points of individual grains. Red line represents a guide to the eye, obtained by a linear fit to the data for the 200 largest grains, giving a distribution $p(k)\sim k^{-\alpha}$ with $\alpha=3.64 \pm 0.02$.}
	\label{Fig3}
\end{figure}

A further interesting property is that the grain-size distribution reveals power-law characteristics, suggesting the existence of scale-invariant features in the polycrystalline IS. This is illustrated in Fig.~\ref{Fig3}, which depicts a log-log representation of the rank-size distribution~\cite{Newman2005,Clauset2009} of the grain sizes obtained for a $10^8$-particle cell, such that the largest and smallest grains are ranked first and last, respectively.  To identify the individual grains and determine their sizes we employed the recently developed grain-segmentation algorithm (GSA) in \verb|Ovito|.~\cite{Stukowski2010a,Larsen2016a} The rank-size graph in Fig.~\ref{Fig3} shows a manifest linear regime for grain sizes $\gtrsim 10^4$ particles, indicating that the grain-size distribution follows a power law asymptotically, indicating scale invariance.~\cite{Newman2005}

\begin{figure} [t!]
	\includegraphics[width=9 cm ]{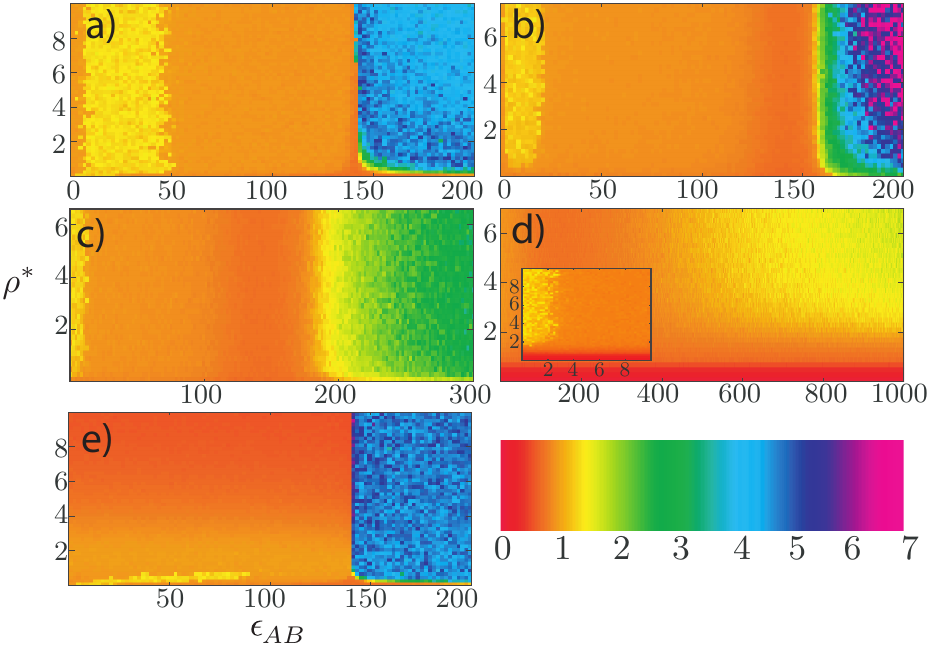}
	\caption{(Color online) Density plots of mean particle displacement in units of the mean interparticle distance $d\equiv{\rho^*}^{-1/3}$ during CG quench as a function of the interaction energy scale $\epsilon_{AB}$ and the reduced density $\rho^*$ for the UF model (a), the IPL4 (b), IPL6 (c), WCA (d) and GC (e) potentials. Inset in (d) shows zoom into region with $\epsilon_{AB}<10\epsilon$. Colors defined in the color bar distinguish between different displacement magnitudes.}
	\label{Fig4}
\end{figure}

In all of the cases shown above, the results are independent of the random initial condition, displaying the same unmixing and crystallization processes for different random-number seeds. Accordingly, for a given particle-number density, the type of  process that occurs is determined by the magnitude of the interspecies interaction strength $\epsilon_{AB}$ only. To further analyze its role we carry out a series of quench CG simulations for a set of $\epsilon_{AB}$-values between 0 and $200\,\epsilon$, employing cells containing of the order of $10^3-10^4$ particles. In addition, we also investigate the possible influence of the particle-number density by considering a range of $\rho^*$-values for each $\epsilon_{AB}$. To automate the detection of the unmixing and crystallization processes we monitor the displacements of the particles during each quench simulation, comparing their positions in the initially structureless state to those at the end of the CG minimization procedure. Fig.~\ref{Fig4}a) displays a density plot of the mean particle displacements (MPD) for the UF system as a function of $\epsilon_{AB}$ and $\rho^*$, expressed in units of the particle-density length scale $d\equiv \rho^{*^{-1/3}}$. It displays three well-defined regimes, characterized by distinct values for the mean particle displacement. The yellow band on the left corresponds to values of the order of $\sim 2d$ and signals the  decay of the initial ideal-gas configuration into the self-similar rock-salt structure.  The mostly blue band on the right corresponds to the unmixing transition in which particles move over significantly larger distances. Finally, in the orange-colored areas the displacements are less than the average particle separation, meaning that the initial configurations are metastable, i.e., they are ``close'' to their corresponding local minima, which retain their chemically uniform and structurally disordered character. A further notable characteristic is that the identification of these 3 groups involves $\epsilon_{AB}$ only, being essentially independent of $\rho^*$, except for very low values for which the distances between the particles become large and the interactions between them weak. This implies that the  ISs associated with high-temperature configurations are invariant with respect to uniform volume scaling.~\cite{Stillinger2015}

Interpreted from the perspective of the PEL formalism,~\cite{Stillinger2015} the above findings imply that, for the considered binary UF model, the topography of the inherent structures for uniformly sampled ideal-gas configurations undergoes abrupt transitions as a function of the interspecies interaction intensity. At $\epsilon_{AB} \simeq 5$ and 50 there is an abrupt transition between chemically uniform, amorphous inherent structures and local minima that display polycrystalline structural order at a homogeneous composition. When reaching $\epsilon_{AB}\simeq 150$, on the other hand, there is a second kind of transition, with the nature of the inherent structures changing from chemically uniform and structurally ordered to compositionally unmixed without long-range structural order.

Another important finding is that the observed phenomena are not limited to the binary UF system but seem to be universal for repulsive interaction potential-energy functions in general. This is illustrated in Figs.~\ref{Fig4}~b-e), which depict density plots of the mean particle distance for the inverse fourth-power law (IPL4), the inverse sixth-power law (IPL6), the Weeks-Chandler-Andersen (WCA) and the Gaussian core (GC) models described in Sec.~\ref{models}, respectively. For all these systems the same 3 regimes can be identified, observing unmixing for large values for $\epsilon_{AB}$,  structural ordering to chemically uniform, rock-salt-type polycrystals for weak interspecies interactions and chemically/structurally amorphous configurations in between. 

A particularly interesting issue in this context concerns the relation between the  crystallization regime and the functional form of the repulsive interaction. Specifically, the shape and the extent of the  structural ordering region in Fig.~\ref{Fig4} is seen to correlate with the rate at which the potential-energy function diverges at the origin. Along the sequence shown in Fig.~\ref{Fig4}~a) to d), in which the divergence changes from slow (logarithmic) to fast ($r^{-12}$), as displayed in Fig.~\ref{Fig1}, the range of energy scales $\epsilon_{AB}$ for which  crystallization occurs reduces systematically. Indeed, the role of the behavior of the pair potential at the origin in the ordering  process becomes even more evident when considering the GC force field, which does not diverge at all, tending to a constant value and zero derivative at the origin. As shown in Fig.~\ref{Fig4}~e), the decay to the rock-salt polycrystal structure in this case is restricted to a very narrow region in the  $\epsilon_{AB}-\rho^*$ plane, disappearing altogether for densities above $\sim 0.8$.

Finally,  the observed unmixing and crystallization phenomena closely resemble the two types of continuous non-equilibrium phase transition that are known to occur in mixtures such as metallic alloys.~\cite{Balluffi2005,Henkel2011}  As discussed in detail in Ref.~\onlinecite{Balluffi2005} such transitions start from an initial instability, meaning that any infinitesimal variation in, for instance, the chemical concentration or degree of crystallinity, lowers the free energy of the system, with spinodal decomposition and a number of structural order-disorder transitions as examples. The phenomena observed here are very similar to such behavior, with the spontaneous decay of infinite-temperature ideal-gas configurations into chemically or structurally ordered states, respectively.
However, a formal classification of these phenomena in terms of such continuous nonequilibrium phase transitions also requires an analysis of the system's time-evolution, which, for such phenomena, is known to exhibit dynamical scaling properties.~\cite{Henkel2011,Laradji1996,Thakre2008} Such an analysis requires modeling protocols that include system dynamics, which is inaccessible to the employed CG minimization protocol here.  

\section{Conclusions}

In conclusion, we have considered  nonequilibrium behavior of classical, 3-dimensional binary mixtures of particles interacting through purely repulsive forces during processes in which an infinite-temperature, ideal-gas initial structure is instantly quenched to zero temperature using conjugate-gradient minimization. We find that such systems display two different types of ordering process which can be controlled by tuning the interactions between unlike particles.  Whereas strong inter-species repulsion giving rise to unmixing, weak interactions lead to a spontaneous development of structural order, forming a rock-salt-type polycrystalline solid of uniform composition. Furthermore, the findings suggest that the dual-type transition behavior is universal for repulsive pair interaction potential-energy functions in general, with the propensity for the crystallization processes being related to their behavior in the neighborhood of zero separation. 
Finally, the observed phenomenology displays features that resemble the two kinds of continuous nonequilibrium phase transitions that are known to occur in mixtures such as metallic alloys. However, a formal classification of these phenomena in terms of such continuous nonequilibrium phase transitions also requires an analysis of the system's time-evolution, which is inaccessible to the minimization protocol employed here.

\section*{Acknowledgments}

We gratefully acknowledge support from the Brazilian agencies CNPq, Capes, Fapesp 2016/23891-6 and the Center for Computing in Engineering \& Sciences - Fapesp/Cepid no. 2013/08293-7. Part of the calculations were performed at CCJDR-IFGW-UNICAMP. The authors acknowledge the National Laboratory for Scientific Computing (LNCC/MCTI, Brazil) for providing HPC resources of the SDumont supercomputer, which have contributed to the research results reported in this paper. URL: http://sdumont.lncc.br. We thank Alexander Stukowksi and Peter Larsen for their assistance with \verb|Ovito|'s grain segmentation algorithm.

\section*{Data availability statement}
The data that support the findings of this study are available from the corresponding author upon reasonable request.

\bibliographystyle{apsrev4-1}

%

\end{document}